\documentclass{article}
\usepackage{spconf,amsmath,graphicx}
\usepackage{thmtools}
\usepackage{thm-restate}
\usepackage{amssymb}
\usepackage[noend]{algpseudocode}
\usepackage{algorithm}

\usepackage[normalem]{ulem}
\usepackage{bbm}
\usepackage{booktabs,multirow}
\usepackage{diagbox}
\usepackage{graphicx}
\usepackage{subfig}
\usepackage{hyperref}
\usepackage[nameinlink]{cleveref}
\usepackage{mwe}
\usepackage[backend=bibtex, maxbibnames=1, sorting=none]{biblatex}
\usepackage[bb=boondox]{mathalfa}

\newtheorem{lemma}{Lemma}

\newif\iflong
\longtrue
\newcommand{\inExtendedVersion}[1]{\iflong #1\fi}

\def\concat{{\mathbin \Vert}}
\def\Concat{{\mathbin \bigg\Vert}}

\title{A Wasserstein graph distance based on distributions of probabilistic node embeddings}

\names{Michael Scholkemper\textsuperscript{1,*} \quad Damin Kühn\textsuperscript{1,*} \quad Gerion Nabbefeld\textsuperscript{2} \quad Simon Musall\textsuperscript{2}  \thanks{* Both authors contributed equally to this research. \\ This work was funded as part of the Graphs4Patients Consortia by the BMBF (Bundesministerium für Bildung und Forschung). We acknowledge partial funding from the DFG RTG 2236 “UnRAVeL” – Uncertainty and Randomness in Algorithms, Verification and Logic." and the Ministry of Culture and Science of North Rhine-Westphalia (NRW Rückkehrprogramm). © 2023 IEEE. Personal use of this material is permitted. Permission from IEEE must be obtained for all other uses, in any current or future media, including reprinting/republishing this material for advertising or promotional purposes, creating new collective works, for resale or redistribution to servers or lists, or reuse of any copyrighted component of this work in other works.}}{Björn Kampa\textsuperscript{2} \quad Michael T. Schaub\textsuperscript{1}}
\address{Department of Computer Science\textsuperscript{1}, Department of Biology\textsuperscript{2}\\ RWTH Aachen University, Germany }
\begin{document}
\ninept%
\maketitle
\begin{abstract}
Distance measures between graphs are important primitives for a variety of learning tasks.
In this work, we describe an unsupervised, optimal transport based approach to define a distance between graphs.
Our idea is to derive representations of graphs as Gaussian mixture models, fitted to distributions of sampled node embeddings over the same space.
The Wasserstein distance between these Gaussian mixture distributions then yields an interpretable and easily computable distance measure, which can further be tailored for the comparison at hand by choosing appropriate embeddings.
We propose two embeddings for this framework and show that under certain assumptions about the shape of the resulting Gaussian mixture components, further computational improvements of this Wasserstein distance can be achieved. 
An empirical validation of our findings on synthetic data and real-world Functional Brain Connectivity networks shows promising performance compared to existing embedding methods.
\end{abstract}

\everymath{\medmuskip=2.5mu minus 2.5mu\thickmuskip=2.5mu minus 2.5mu}

\section{Introduction}
\label{sec:intro}
Graphs and networks have become an almost ubiquitous abstraction in domains like biology, medicine or social sciences to represent a large range of complex systems~\cite{strogatz2001exploring}. 
For instance, protein interactions, brain connections or social dynamics are frequently modelled as networks and studied from this perspective~\cite{szklarczyk2015string,rubinov2010complex,borgatti2009network}.
Due to this increasing abundance of network data, the classical problem of quantifying (dis-)similarities between graphs has seen a surge of research interest recently. %
Indeed, a graph distance measure to compare the structure of various systems is crucial to enable an exploratory, comparative analysis of (sets of) graphs in many application contexts.
However, for most applications it is typically not only important to quantify the difference between two graphs on a global level, but to identify the lower-level, structural differences that contribute to this difference. 
Accordingly, optimal transport based graph distances, which not only provide a distance measure between two graphs based on a probabilistic matching but also a transport plan that highlights where changes occur, have recently gained significant attention~\cite{petric2019got,barbe2020graph,maretic2022fgot}.%

In general, graph similarity measures may be classified as either supervised or unsupervised.
Supervised approaches aim at learning a distance function that effectively distinguishes between differently labeled networks.
These include approaches for graph similarity of human brain fMRI data using Graph Neural Networks~\cite{ma2019deep} or Protein-Protein interactions using Genetic Programming~\cite{sousa2020evolving}. 
Unsupervised approaches, on the other hand, are concerned with finding distances between networks without having access to labels. 
They are particularly useful for the exploratory study of cluster differences beyond known classes. 
Some approaches leverage the powerful but computationally expensive Graph Edit Distance~\cite{mohapatra2022chemistry,gao2010survey}.
Other methods first compute a vector representation of the network, which is then used to define a distance metric~\cite{mo2022simple, riba2021learning}.
Recently, there have been approaches that use Optimal Transport (OT) to define a distance between networks, based on the Gromov-Wasserstein distance~\cite{memoli2011gromov,peyre2016gromov}.%
Fused Gromov-Wasserstein~\cite{titouan2019optimal} is an extension for attributed graphs where in addition to the graph structure, node attributes can also influence the distance between two graphs. 
Closely related to our approach are OT-based methods that leverage Wasserstein distances on graphs~\cite{petric2019got,barbe2020graph,maretic2022fgot}. 
These approaches define the distance between two graphs as the distance between the distributions of the corresponding systems excited by Gaussian noise. 
In contrast, we propose specific non-gaussian node embeddings that highlight distinct structural aspects of the graph.

\noindent\textbf{Contribution.}
In this paper, we propose a novel unsupervised approach for computing the distance between two graphs based on Optimal Transport. 
This provides us not only with an alignment between the two node sets of the graph, but also with a measure of the quality of this alignment (the actual distance between the graphs).
Our approach is efficient and thus scalable to large data sets.
Further it can even be used to compare graphs of different sizes.
We show that, as we increase the number of samples, our approach defines a distance pseudometric on the space of all graphs.
Further, we evaluate our approach on a range of synthetic data and apply it to Functional Brain Connectivity networks of mice, where we can recover meaningful patterns in the data.

\section{Notation and Preliminaries}\label{sec:notation}
\noindent\textbf{Notation.}
A graph $G = (V, E)$ consists of a node set $V$ and an edge set $E = \{uv \mid u,v \in V\}$.
Given a graph $G = (V,E)$, we identify the node set $V$ with $\{1, \ldots, n\}$. 
We allow for self-loops $vv \in E$ and positive edge weights $w: E \rightarrow \mathbb{R}_+$ in our graphs.
For a matrix $M$, $M_{i,j}$ is the component in the $i$-th row and $j$-th column. 
We use $M_{i,\_}$ to denote the $i$-th row vector of $M$ and $M_{\_,j}$ to denote the $j$-th column vector.
An \emph{adjacency matrix} of a given graph is a matrix $A$ with entries $A_{u,v} = 0$ if $uv \notin E$ and $A_{u,v} = w(uv)$ otherwise, where we set $w(uv)=1$ for unweighted graphs for all $uv\in E$.
For two vectors $x, y$ we write $x \concat y$ for the concatenation and $\concat_{i=0}^n x_i$ for the concatenation over a sequence of vectors $x_0, \ldots, x_n$.  
We denote the two norm of a vector $x$ by $\|x\|$.

\noindent\textbf{Optimal Transport.} Optimal transport (OT) is a framework for computing distances between probability distributions. 
In this paper, we leverage the so called \textit{Wasserstein distance} ($W_2^2$), also referred to as Earth Movers Distance. 
For two probability distributions $\mathcal{X},\mathcal{Y}$ on some metric space $\mathcal{S}$, the Wasserstein distance can be computed by solving the following optimization problem:
\begin{equation}
    \begin{aligned}
    \label{def:W2}
    W_2^2(\mathcal{X},\mathcal{Y}) = \min_{\pi \in \Pi(\mathcal{X},\mathcal{Y})} \int_{\mathcal{X} \times  \mathcal{Y}} ||x - y||^2 d\pi(x,y) \\ %
    \end{aligned}
\end{equation} %
where $\Pi(\mathcal{X},\mathcal{Y})$ is the set of all admissible couplings $\pi$ on $\mathcal{S} \times \mathcal{S}$ whose marginals are $\mathcal{X}$ and $\mathcal{Y}$ with $\pi(x,y)$ being the mass moved from $x$ to $y$.

When both distributions considered are multivariate Normal distributions, i.e., $\mathcal{X} = \mathcal{N}(\mu_1, \Sigma_1), \mathcal{Y}   = \mathcal{N}(\mu_2, \Sigma_2)$, with mean vectors $\mu_1,\mu_2$ and covariance matrices $\Sigma_1,\Sigma_2$, respectively, the Wasserstein distance has a closed form expression given by
\begin{equation*}
    W_2^2(\mathcal{X},\mathcal{Y}) = ||\mu_1 - \mu_2|| + \operatorname{tr}(\Sigma_1) + \operatorname{tr}(\Sigma_2) 
    - 2\operatorname{tr}((\Sigma_1^{\frac{1}{2}}\Sigma_2\Sigma_1^{\frac{1}{2}})^{\frac{1}{2}})
\end{equation*}
Reference~\cite{delon2020wasserstein} appropriately generalizes this to Gaussian Mixtures $\mathcal{M}, \hat{\mathcal{M}}$, which are central to our approach. Here we consider uniformly weighted GMMs $\mathcal{M}= \frac{1}{n} (\mathcal{N}_1 + ... + \mathcal{N}_n)$ with Gaussian distributions $\mathcal{N}_i = \mathcal{N}(\mu_i, \Sigma_i)$, called Gaussian components, having equal weight $\frac{1}{n}$.
In this case the optimal transport distance can be computed by considering the optimization problem
\begin{equation}
    \label{def:MW2}
    MW_2^2(\mathcal{M},\hat{\mathcal{M}}) = \min_{\pi \in \Pi(\mathcal{M},\hat{\mathcal{M}})}\sum_{i,j} W_2^2(\mathcal{N}_i,\hat{\mathcal{N}}_j) \pi_{i,j}
\end{equation}
where $\mathcal{N}_i$, $\hat{\mathcal{N}}_j$ are the $i$-th resp. $j$-th component of the Gaussian mixture distributions $\mathcal{M},\hat{\mathcal{M}}$ and $\pi_{i,j}$ the mass moved from the Gaussian component $\mathcal{N}_i$ to the Gaussian component $\hat{\mathcal{N}}_j$. 

The OT framework can also be used to compare metric spaces (or distributions of points defined in different spaces) by means of the \textit{Gromov-Wasserstein distance} ($GW_2^2$) \cite{memoli2011gromov,peyre2016gromov}. 
For two probability distributions $\mathcal{X},\mathcal{Y}$ supported in different spaces the associated optimization problem the becomes~\cite{salmona2021gromov}:
\begin{equation*}
    \label{def:GW2}
    GW_2^2(\mathcal{X},\mathcal{Y}) = \min_{\pi \in \Pi(\mathcal{X},\mathcal{Y})} \sum || W_2^2(x_i,x_k) - W_2^2(y_j,y_l)||^2\pi_{i,j}\pi_{k,l}
\end{equation*}
where $x_i,x_k \sim \mathcal{X} $ and $y_j,y_l \sim \mathcal{Y}$.

Intuitively, the Gromov Wasserstein formulation seeks to map points onto each other such that the overall distances between all pairs of points are as much as possible preserved.
Hence, in contrast to \autoref{def:MW2} the Wasserstein distances $W_2^2$ are only computed between elements of the respective distributions $\mathcal{X}$ and $\mathcal{Y}$. 
Consequently the Gromow Wasserstein distance $GW_2^2$ can be computed for distributions supported on different spaces.
While this additional flexibility can be advantageous for applications, in this paper we argue that it can be worthwhile to compute vectorial graph representations that are supported in the same space.
This enables us to leverage the Wasserstein Distance as a graph distance measure. Our experiments show that our proposed embedding methods CCB and CNP are suited to produce such graph representations.

\section{Proposed Method}
\label{sec:method}

In this section, we establish our approach for computing the distance between two graphs using OT. 
The high-level approach is as follows: 
We compute multiple randomly initialised i.i.d node embedding for each node. 
Subsequently fitting a Gaussian to the sampled embeddings of each node represents the graph as a Gaussian Mixture. By computing the optimal transport plan between the Gaussian Mixtures of two graphs we obtain a node allignment with the corresponding cost. 
In the following, we present two node embeddings that can be used in the above framework and that highlight different properties of the network. 

\noindent\textbf{Node Embedding.} Our approach hinges on the fact that the embedding we create for each node is dependent on some random variable. If this is not the case, then the (co-)variance is jointly $0$ for all nodes which reduces the Wasserstein distance to the square euclidean distance between the means.
We propose two node embeddings that fulfill this requirement: CCB and CNP. 
The proposed \textit{Colored Cooper Barahona} embedding (CCB) is an extension of the \textit{Cooper Barahona} embedding \cite{cooper2010role}.
The original embedding embeds a node as the concatenation of the rows in the matrix power $A^{\delta}\mathbb{1}$ of the adjacency matrix $A$. 
This captures not only the degree of a node but also the connections of length up to $\delta < d$.
We adapt the embedding by using colors which we use to combine the nodes into groups. 
We thus receive a more expressive yet still low-dimensional embedding.

The CCB embedding works as follows: 
For a number of colors $k$ and $n$ nodes, we sample $k-1$ cuts $(c_2, ..., c_{k})$ uniformly at random from $\{1,..., n-1\}$ without replacement, sort them such that $c_i < c_j$ for $i < j$, and define $c_1 = 0, c_{k+1} = n$. We then construct a block matrix $H \in \{0,1\}^{k \times n}$ where $H_{i,j} = 1$ if $c_j \leq i \leq c_{j+1}$ and $H_{i,j} = 0$ otherwise for $1 \leq j \leq k$.
We then simply compute the embedding as:
\begin{equation*}
    \bar{\varphi}_{\text{CCB}}(v, H, d) = \Concat_{i=0}^d \frac{1}{\|A\|^i}A^i_{v, \_} H.
\end{equation*}
The CCB embedding thus embeds the nodes with an embedding of size $k \cdot d$. However, the ordering of the nodes is paramount for the expressivity of the embedding. A new ordering of the nodes could lead to vastly different embeddings.

On the contrary, the \textit{Colored Neighborhood Propagation} (CNP) embedding is one that is invariant under reordering of the nodes.
It is an extension of the SNP embedding used in \cite{scholkemper2021local}. 
Again, we adapt the original using random colors, the sampling procedure is, however, different: 
We first randomly assign one of $k$ colors to each node using an indicator matrix $H$, where $H_{v, c} = 1 \iff c(v) = c$ and $H_{v, c} = 0$ otherwise. As opposed to the CCB embedding, this matrix has no block structure. 
For each distance $0 \leq \delta \leq d$ and for each node $v$, we count the number of nodes reachable in $\delta$ steps, which have a certain color, and store them in a matrix of size $(d+1) \cdot k$. 
We then sort the colums of this matrix lexicographically. This leads to the following definition of the CNP embedding:
\begin{equation*}
    \bar{\varphi}_{\text{CNP}}(v, H, d) = \Concat_{i=1}^{d+1} M_{i,\_} \text{ with }
    M =\text{lex-sort}\left( \begin{bmatrix}
        \frac{1}{\|A\|}A^0_{v, \_} H\\
        \vdots \\
        \frac{1}{\|A\|^d}A^d_{v, \_} H
        \end{bmatrix}\right)
\end{equation*}
We also normalize each embedding $\varphi(v, H, d) = \frac{1}{||\bar{\varphi}(v, H, d)||}\bar{\varphi}(v, H, d)$.

Due to the sorting, this embedding is invariant under isomorphism, that is, the probability of sampling a certain embedding is independent of the node ordering of the graph. 

\begin{algorithm}[t]
    \begin{algorithmic}[1]
        \State \textbf{Input:} $G_1=(V_1, E_1), G_2=(V_2, E_2)$
        \For{$v \in V_1 \cup V_2$}
            \For{$i \leq s$}
                \State sample assignment of node to colors $H^{(i)}$ 
                \State $\varphi^{(i)}(v) = \varphi_{\text{X}}(v, H^{(i)}, d)$
            \EndFor
            \State Fit Gaussian $\mathcal{N}(\mu_v, \Sigma_v)$ on $\varphi^{(1)}(v), ..., \varphi^{(s)}(v)$
        \EndFor
        \State Compute Gaussian Mixture $\mathcal{M}_x = \sum_{v \in V_x}\mathcal{N}(\mu_v, \Sigma_v)$
        \Return $MW_2^2(\mathcal{M}_1, \mathcal{M}_2)$
    \end{algorithmic}
    \caption{Compute the distance between $G_1$ and $G_2$}
    \label{alg:1}
\end{algorithm}

\noindent\textbf{Optimal Transport of Gaussian Mixtures.} For each node $v$ we compute $s$ embeddings $\varphi^{(1)}, ..., \varphi^{(s)}$. 
We now fit a Gaussian using the maximum likelihood estimate $\mathcal{N}_v = \mathcal{N}(\hat{\mu}, \frac{1}{n}\sum_i (x_i -\hat{\mu})(x_i -\hat{\mu})^\top)$ on this collection $\{.., x_i, ..\}$ of embedding points, where $\hat{\mu} = \frac{1}{n}\sum_i x_i$.
The entire graph is then encoded as a Gaussian Mixture $\mathcal{M}(G) = \frac{1}{|V|}\sum_{v \in V}\mathcal N_v$ of the Gaussians extracted from each node. 
To compute the distance between two graphs, we can then compute the Wasserstein distance between the two Gaussian Mixtures $\mathcal{M}(G_1), \mathcal{M}(G_2)$ (see \cref{def:MW2}).
The whole procedure can be found in \Cref{alg:1}.
This leads to relevant properties of the extracted distances for both CCB and CNP embeddings: 
\begin{restatable}{proposition}{CNPPSEUDOMETRIC}
    \label{prop:cnp_pseudometric}
    For the sample size $s \rightarrow \infty$, CNP defines a pseudometric on the space of all graphs and CCB defines a pseudometric on the space of all adjacency matrices.
\end{restatable}
The distinction here is related to the isomorphism invariance of the embeddings. While CNP converges to the same expectation regardless of the node ordering, CCB is dependent on the node ordering and will assign a non-zero distance to isomorphic graphs.

The following proposition states that our distance measures can be simplified if we assume additional conditions on the covariances of the distributions.
\begin{restatable}{proposition}{WASSERSTEINTHREE}
    \label{thm:minithree}
    Let $D_i = \operatorname{diag}(d^{i}_1, ..., d^{i}_n), D_j = \operatorname{diag}(d^{j}_1, ..., d^{j}_n)$.
    Assume the Mixture components $\mathcal{N}_i = \mathcal{N}(\mu_i, \Sigma_i)$ share scaled covariances: $D_i\Sigma_i  = D_j \Sigma_j = \Sigma$. Let $\lambda_x$ be the eigenvalues of $\Sigma$. Then, the Wasserstein distance between two components is equal to:
    \begin{equation*}
        \begin{aligned}
            W_2(\mathcal{N}^{G}_i, \mathcal{N}^{\hat{G}}_j) &= \|\mu_i - \mu_j\|_2^2 + \sum_{x=1}^n \frac{\lambda_x}{d^{i}_x} +  \frac{\lambda_x}{d^{j}_x} - \frac{2\lambda_x}{\sqrt{d^{i}_x d^{j}_x}}
        \end{aligned}
    \end{equation*}
\end{restatable}
This substantially speeds up the computation as we do not have to compute a matrix square root.
While the assumptions on the covariance may not always be fulfilled, we can use of the above formula as an approximation.
In the following, we use three different approaches to compute the distance between two Gaussians that only differ in the approximation of the Wasserstein distance used: 
The \textit{full} Wasserstein distance, the \textit{scaled} Wasserstein distance, where we adjust the covariances of the Gaussian components such that the assumption of \Cref{thm:minithree} holds, and the \textit{tied} Wasserstein distance, where we assume $\Sigma_i = \Sigma_j = \Sigma$, which further simplifies the Wasserstein distance to only the square euclidean distance between the means.

\noindent\textbf{Properties of our approach.}
We remark that with our computations, we not only obtain a distance between two graphs, but also a (probabilistic) mapping between the nodes via the computed transport plan. 
For applications, this alignment can be very useful. 
Furthermore, one can use OT to compute the distance between two Mixtures with a distinct number of components --- meaning that we can compare graphs of different sizes.
One can even define unbalanced transport plans, such that only similar nodes are mapped to each other.
Moreover, our approach using the tied Wasserstein distance is very efficient making it applicable to (sparse) graphs of size $|V| \approx 10.000$. 
For even larger graphs, reducing the number of Mixture components \cite{crouse2011look,assa2018wasserstein} can be used to speed up computations even further.

\section{Evaluation}
\label{sec:experiments}

We now present two experiments on both synthetic and real-world biological data to evaluate the performance of our OT-based graph distances in comparison to other common distance measures.
First, we qualitatively show the meaningfulness of clusters produced from our distance measures and the capability to retain these clusters under 2D projections, commonly used in biological domains to visualize population differences. 
Second, a classification task is presented to provide a quantitative comparison of our approaches with other graph distance methods. 
It should be noted that although a supervised classification task is presented, all methods are unsupervised and do not use the labels in any way apart for the evaluation.
We benchmark against the Euclidean distance between the degree (Degree) distributions, the dominant eigenvector (EV) and the Graph2Vec \cite{narayanan2017graph2vec} embedding of the graphs. We also compare against the Node2Vec \cite{grover2016node2vec} and Role2Vec \cite{ahmed2019role2vec} embeddings using Gromov-Wasserstein as a distance measure, as well as the GOT \cite{petric2019got} distance.
All code and data used for the experiments are available \href{https://git.rwth-aachen.de/netsci/wasserstein-graph-dist-prob-embeddings/}{here}\footnote{git.rwth-aachen.de/netsci/wasserstein-graph-dist-prob-embeddings/}.

\everymath{\medmuskip=2.5mu minus 2.5mu\thickmuskip=2.5mu minus 2.5mu}
\noindent\textbf{Synthetic Networks.} 
This dataset consists of random networks generated with four common network models: Erd\H{o}s-Rényi (ER), Watts-Strogatz (WS), Barabási-Albert (BA) and Configuration Models (CF) \cite{erdHos1960evolution,watts1998collective,barabasi1999emergence,newman2001random}. 
From each of the four generative models we sample $20$ networks with $n \in \{10,200\}$ nodes. 
The other parameters such as edge probability or degree distribution are chosen such that nodes in the resulting networks have an expected degree of 6. 

\noindent\textbf{Functional Brain Connectivity Networks.} 
This dataset consists of Functional Brain Connectivity (FC) networks \cite{biswal1995functional,friston2011functional} calculated as the Pearson-Correlation between the neural activity traces of different brain regions defined by the Allen Brain Atlas \cite{sunkin2012allen}.
This results in complete, weighted graphs with $n=64$ nodes. 
The neural activity is recorded through Widefield Calcium Imaging \cite{homma2009wide,scott2018imaging,cramer2019vivo,wekselblatt2016large} while the mice perform a virtual maze experiment under a two-alternative forced choice paradigm\cite{leinweber2014two,pinto2018accumulation,mayrhofer2013novel,scott2015sources}.
A trial in this experiment consists of the mice perceiving a uni- or multisensory stimulus and responding accordingly after a short delay to get rewarded.
The FC networks we use are grouped into 5 classes: Default Mode Network (DMN), Left- \& Right Stimulus (LS\&RS) and Left- \& Right Response (LR\&RR). While the DMN corresponds to the baseline neural connectivity between the trials, the stimuli and response classes contain the FC networks from the respective phases within the trial. Each class contains 200 FC networks from 3 different experimental sessions of the same subject.

\newlength{\tempdima}
\newcommand{\rowname}[1]%
{\rotatebox{90}{\makebox[\tempdima][c]{\textbf{#1}}}}
\newcommand{\subfigimg}[3][,]{%
  \setbox1=\hbox{\includegraphics[#1]{#3}}%
  \leavevmode\rlap{\usebox1}%
  \rlap{\hspace*{10pt}\raisebox{\dimexpr\ht1-10\baselineskip}{#2}}%
  \phantom{\usebox1}%
}

\begin{figure}[t]
    \centering
    \begin{tabular}{@{}c@{}c@{}c@{}}
        \subfloat[Synthetic Networks Distances]{\includegraphics[trim=0cm 0.1cm 0.5cm 0.2cm, clip, width=0.27\textwidth]{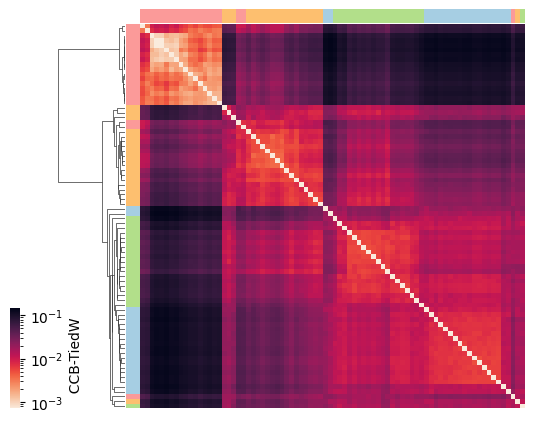} } &
        \subfloat[CCB-TiedW UMAP Projection]{\includegraphics[trim=1cm 1.1cm 0.25cm 0.82cm, clip, width=0.23\textwidth]{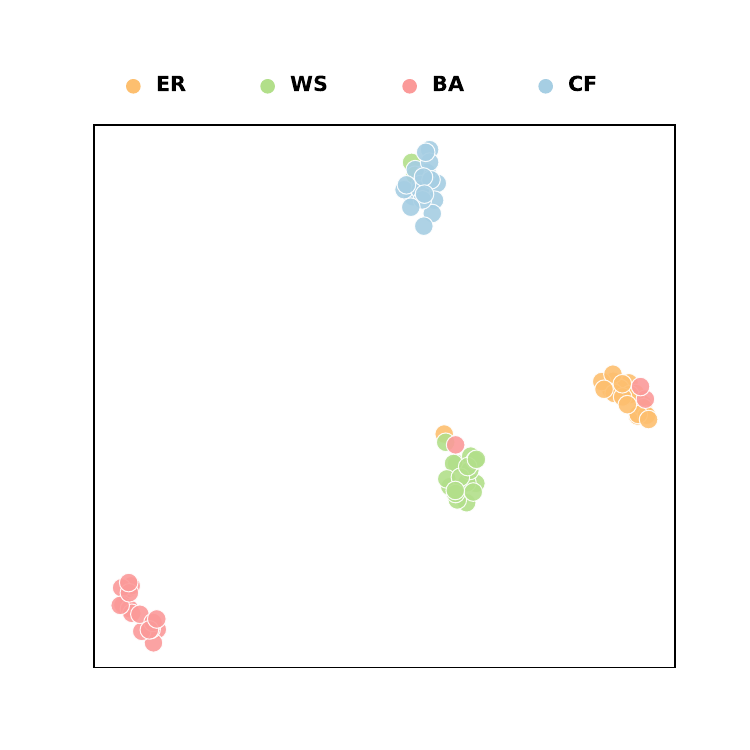}} \\[-2 ex]
        \subfloat[FC Network Distances]{\includegraphics[trim=0cm 0cm 0.5cm 0.2cm, clip, width=0.27\textwidth]{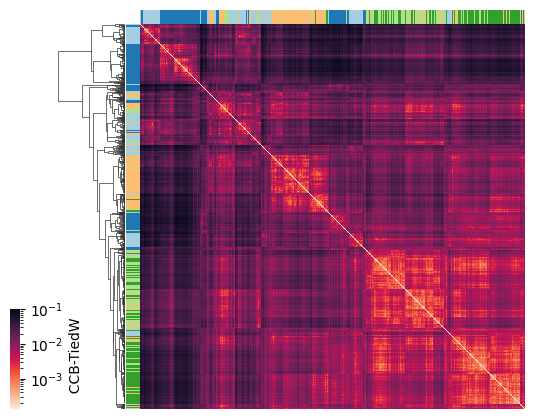}}  &
        \subfloat[CCB-TiedW UMAP Projection]{\includegraphics[trim=1cm 1.1cm 0.25cm 0.82cm, clip, width=0.23\textwidth]{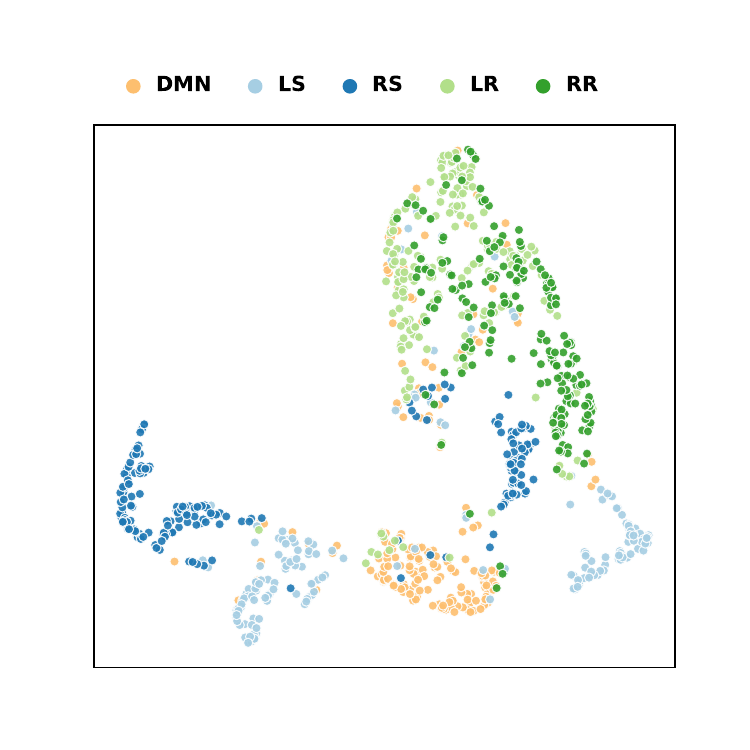}} \\ [-1 ex]
    \end{tabular}
    \caption{CCB-TiedW distances for the synthetic networks (a) and the functional connectivity networks (c). The networks are ordered according to the hierarchical clustering dendrogram where small heights correspond to small cluster distances. These distances can also be projected into a 2D space using UMAP for visualization purposes (b,d). Class memberships are indicated by the same color scheme in both corresponding plots.}
    \label{fig:1}
\end{figure}

\noindent\textbf{Setup.}
On both the synthetic and the real-world dataset, we compute the pairwise distances as defined by CCB and CNP between all graphs. For both experiments the chosen parameters for our embedding methods on the real world data were sampled $s=1000$ times with $k = 10$ colors and depth $d=5$. 
Larger values for these parameters generally did not decrease performance, but only increased computation times. 
Both embeddings therefore do not appear sensitive to the specific parameter selection.
We aim to show this more rigorously as part of a sensitivity analysis in future work. 
To ensure a fair comparision, all competing node embedding methods, like Node2Vec and Role2Vec were computed for the same number of total dimensions, while Graph2Vec was allowed larger dimensions as it computed only one embedding per graph. 
The resulting distance matrices are depicted as a hierarchically clustered heatmap in \autoref{fig:1}. 
In the same figure, we also show a 2D projection of the distance landscape using UMAP \cite{mcinnes2018umap}.
As a qualitative comparison, a $k$-Nearest Neighbor (kNN) classification \cite{cover1967nearest} is performed on both datasets based on the precomputed distances. 
This provides a measure for how proximities in these distances reflect the true class membership of the graphs. 
For this purpose, a weighted kNN classifier ($k=5$) is used which weights points by the inverse of their distance. This gives a higher importance to closer neighbors. 
To validate the generalizability of the computed distances a k-Fold Cross-validation scheme is deployed. 
This means that the neighbors the classification is based on are from a training subset of the graphs while the evaluation of the actual classification is done on a separate test set containing unseen graphs. 
We test on $20\%$ of the data in each of the 20 splits.
The mean accuracy and its standard deviation over the splits are shown in \autoref{tab:1}. 
We also report the silhouette score as a measure of cluster density. Computation times are given as an average over all pairwise computed distances. \\
\noindent\textbf{Discussion.}
On the synthetic graphs we can see that the clusters are generally well separated with small inner cluster distances and large distances between clusters. 
Additionally, the hierarchical clustering shows that these clusters can be found by relatively simple algorithms given our precomputed distances. 
This stands in stark contrast to the distances computed by other approaches that did not recover any meaningful clusters. \autoref{tab:1} also shows this trend as classification and silhouette scores are high for our approaches and considerably worse for the competition. The corresponding figures for all considered methods can be found in the supplementary material. \\
In the real world data, we can see the presence of noise, with some networks not clearly distinguished according to their classes. However, this is to be expected due to the nature of behavioral experimental data where observed behaviors are not always caused by the activation of the hypothesized neural pathways. More specifically, while we have only included trials where the mouse responded correctly to the presented stimuli, this behavior might also happen by random chance if the mouse is disengaged. Despite the presence of noise, CCB-TiedW successfully finds meaningful clusters w.r.t the defined classes. Most trials from the DMN, LR and RR classes are clearly clustered together in the heatmap and projection in \autoref{fig:1}. Interestingly, trials from the LS and RS both form two well separated sub-clusters with similar structures which can not be explained by the stimulus type as visual and tactile trials were present in both clusters. This could hint at trials where the mouse was not engaged and that are thus further away from the responses and closer to the default mode network. Such exemplary findings of unexpected but consistent within-population differences beyond known labels illustrate how unsupervised methods can help explore real-world data and provide directions for further investigation. \\
The run times of CNP and CCB in \autoref{tab:1} show that the tied and scaled Wasserstein formulations provide a significant speed up without a loss in performance. Competing OT based methods are on average around 3 to 10 times slower, while euclidean distance based methods are faster but fail to differentiate the graph classes.

\begin{table}[t]
    \resizebox{\columnwidth}{!}{%
    \centering
    \footnotesize
    \begin{tabular}{l@{\hspace{0pt}}r|c@{\hspace{4pt}}c@{\hspace{2pt}}cc@{\hspace{4pt}}c@{\hspace{2pt}}c} %
        \toprule
        \multicolumn{2}{l|}{\multirow{2}{*}{\backslashbox{Method}{Dataset}}} & \multicolumn{3}{c|}{Random Graphs}     & \multicolumn{3}{c}{Functional Connectivity}\\
        &                                                                               & KNN  & Silh. & \multicolumn{1}{c|}{t (\textit{ms})}    & KNN & Silh. & t (\textit{ms})      \\
        \midrule
        Degree &                                                                        & 0.25±0.1 & -.082 &  \textless0.01                                    & 0.53±0.04 & -.074         &  \textless0.01 \\
        EV &                                                                            & 0.59±0.09 & .02 &  0.05                                    & 0.44±0.03 & -.047         &  0.01 \\
        \multicolumn{2}{l|}{Graph2Vec}                                                  & 0.51±0.14 & .01 &  0.08                                   & 0.33±0.02 & -.168         &  0.01\\
        \midrule
        Node2Vec   & GW                                                                 & 0.61±0.10 & -.003 &  390                                              & 0.76±0.03	 & \textbf{.133} &  14.74 \\
        \multicolumn{2}{l|}{Role2Vec}                                                   & 0.71±0.10 & -.014 &  109                                             & 0.78±0.03 & .032           &  9.67\\
        \midrule
        GOT  & W                                                                        & -- & -- & --                                                          & 0.68±0.03 & -.209         &  24.30 \\
        \multicolumn{2}{l|}{CNP-Tied}                                                   & 0.90±0.06 & \textbf{.550} &  40                                             & 0.59±0.03 & -.169 &  2.85\\
        \multicolumn{2}{l|}{CCB-Tied}                                                   & 0.91±0.06 & .353 &  36                                              & 0.82±0.02 & -.019         &  2.60\\
        \multicolumn{2}{l|}{CNP-Scaled}                                                 & \textbf{0.93±0.07} & .512 &  57                                              & 0.58±0.03 & -.170&  14.22\\
        \multicolumn{2}{l|}{CCB-Scaled}                                                 & 0.90±0.06 & .385 &  52                                              & 0.81±0.03 & -.021         &  14.11\\
        \multicolumn{2}{l|}{CNP-Full}                                                   & \textbf{0.93±0.05} & .528 &  178                                              & 0.59±0.03 & -.167         &  36.57\\
        \multicolumn{2}{l|}{CCB-Full}                                                   & 0.92±0.05 & .358 &  170                                              & \textbf{0.83±0.02} & -.015&  50.54\\
        \bottomrule
    \end{tabular} }
    \normalsize
    \caption{Weighted k-NN ($k=5$) classification scores on synthetic and real-world data given as $\mu \pm \sigma$. Classification is performed under a 20-fold cross validation with a relative test set size of 20\%. OT-based methods are grouped into Gromov-Wasserstein (GW) and Wasserstein (W) distances. Computation times t are averaged over all pairwise computed distances. -- indicates that the provided method is not implemented for graphs of different sizes.}
    \label{tab:1}
\end{table}

\section{Conclusion}
\label{sec:conclusion}
We introduced an Optimal Transport framework that represents each graph as the Gaussian Mixture of probabilistic node embeddings. 
This enabled the use of the Wasserstein distance instead of the widely used Gromov Wasserstein distance.
We introduced two probabilistic node embeddings that fulfill the requirements of the framework and highlight different properties of the graph. 
Further, we derived theoretical properties of the resulting graph distances showed their efficiency and performance on both synthetic and real-world data.

\vfill\pagebreak

\section{References}
\printbibliography[heading=none]

\inExtendedVersion{
\normalsize    
\appendix
\section{Plots and Tables}

\section{Proofs.}

\CNPPSEUDOMETRIC*
\textit{Proof.} 
Toward proving this claim given a graph $G$, we will first show that the Gaussians Mixture $\mathcal{M}(G)$ (see \cref{alg:1}) converges to a Gaussian Mixture $\overline{\mathcal{M}}(G) := \lim_{s \rightarrow \infty} \mathcal{M}(G)$. For the moment, assume this to be true. Then, we 
can use the following result: 
\begin{lemma}[Proposition 5, \cite{delon2020wasserstein}] The distance $MW^2_2$ between Gaussian Mixtures (\cref{def:MW2}) defines a metric on the space of Gaussian Mixtures.
\end{lemma}
For graphs $G_1, G_2, G_3$, this means that:
\begin{equation}
    \begin{aligned}
    \operatorname{dist}(G_1, G_1) &= MW^2_2(\overline{\mathcal{M}}(G_1), \overline{\mathcal{M}}(G_1)) = 0\\
    \operatorname{dist}(G_1, G_2) &= MW^2_2(\overline{\mathcal{M}}(G_1), \overline{\mathcal{M}}(G_2))\\ 
    &= MW^2_2(\overline{\mathcal{M}}(G_2), \overline{\mathcal{M}}(G_1)) = \operatorname{dist}(G_2, G_1)\\
    \operatorname{dist}(G_1, G_3) &= MW^2_2(\overline{\mathcal{M}}(G_1), \overline{\mathcal{M}}(G_3))\\
    &\leq MW^2_2(\overline{\mathcal{M}}(G_1), \overline{\mathcal{M}}(G_2)) \\
    &\quad \quad + MW^2_2(\overline{\mathcal{M}}(G_2), \overline{\mathcal{M}}(G_3))\\
    &= \operatorname{dist}(G_1, G_2) + \operatorname{dist}(G_2, G_3)
\end{aligned}
\end{equation}
proving that the distance is a pseudometric.
To show convergence, consider the embedding $\varphi^{(i)}(v)$. Each instance is bounded by $||\varphi^{(i)}(v)||_\infty \leq \delta^d$ where $d$ is the number of adjacency matrix powers used and $\delta$ is the maximum degree in the graph. Since we only allow non-negative edge weights, each component is bounded by $0 \leq \varphi^{(i)}_j(v) \leq \delta^d$. We apply Hoeffdings inequality \cite{vershynin2020high}:
\begin{equation*}
    Pr\left(\left| \frac{1}{s} \sum_{i=1}^s \varphi^{(i)}_j(v) - \mathbb{E}[\varphi_j(v)] \right| \geq \epsilon \right) \leq 2 \exp\left( - \frac{2s \epsilon^2}{\delta^{2d}} \right)
\end{equation*}
We now union bound over all $kd$ components of the embedding $\varphi^{(i)}(v)$:
\begin{equation*}
    Pr\left(\left|\left| \frac{1}{s} \sum_{i=1}^s \varphi^{(i)}(v) - \mathbb{E}[\varphi(v)] \right|\right|_\infty \mskip-15mu \geq \epsilon \right) \leq 2kd \exp\left( - \frac{2s \epsilon^2}{\delta^{2d}} \right)
\end{equation*}
This proves, that the maximum likelihood estimator $\frac{1}{s} \sum_{i=1}^s \varphi^{(i)}(v)$ converges to the expectation $\mathbb{E}[\varphi(v)]$ as $s \rightarrow \infty$. 
For the covariance, we apply the matrix Bernstein inequality (Corollary 6.2.1, \cite{tropp2015introduction}) to the our maximum likelihood covariance estimator $\frac{1}{s}\sum_{i=1}^s (\varphi^{(i)} - \mu) (\varphi^{(i)} - \mu)^\top$. Let $x_i = (\varphi^{i}(v) - \frac{1}{s} \sum_{i=1}^s \varphi^{(i)}(v))$, then:
\begin{equation*}
    Pr\mskip-4mu\left(\left|\left| \frac{1}{s} \sum_{i=1}^s x_i x_i^\top \mskip-10mu - \mathbb{E}[x x^\top] \right| \right|_\infty \mskip-15mu \geq \epsilon\mskip-3mu \right)\mskip-5mu \leq 2 kd \exp\mskip-3mu\left( - \frac{s \epsilon^2/2}{kd\delta^2 (\frac{2}{3} + \delta^2)} \right)
\end{equation*}  
Again this proves that the maximum likelihood estimator fo the covariance converges to the expected covariance as $s \rightarrow \infty$. Combining the two results, we can see that the Gaussian component $\mathcal{N}_v$ representing a node in the graph converges to the expected Gaussian $\overline{\mathcal{N}}_v = \mathcal{N}(\mathbb{E}[\varphi(v)], \mathbb{E}[x x^\top])$. One final union bound yields that the whole Gaussian Mixture converges to a Gaussian Mixture $\overline{\mathcal{M}}(G) = \sum_{v \in V} \overline{\mathcal{N}}_v$ as $s\rightarrow \infty$. 
Finally, to show that the CNP is a pseudometric on the space of graphs, we can use the same argument as above. Additionally we need to show that CNP converges to the same Gaussian Mixture for two isomorphic graphs $G \simeq G'$. 
Let $A, A'$ be the adjacency matrices of $G, G'$ respectively, then $P A P^\top = A' $ for some permutation matrix $P$. 
Recall the definition of CNP:
\begin{equation*}
    \bar{\varphi}_{\text{CNP}}(v, H, d) = \Concat_{i=1}^{d+1} M_{i,\_} \text{ with }
    M =\text{lex-sort}\left( \begin{bmatrix}
        \frac{1}{\|A\|}A^0_{v, \_} H\\
        \vdots \\
        \frac{1}{\|A\|^d}A^d_{v, \_} H
        \end{bmatrix}\right)
\end{equation*}
and consider what happens when you permute the rows of $H$ (so that all nodes have the same color in both graphs) and after the transformation, permuting them back:
\begin{equation*}
P^\top A' P H = P^\top PAP^\top P H = A H
\end{equation*}
This also extends to matrix powers. It the node $v$ has the same color as the node $v'$ it is isomorphic to in the other graph, so the Gaussian Mixture will have the exact same components (in a different order). Also, since $H$ is sampled uniformly i.i.d, $H$ and $P H$ have the same probability to be sampled. Thus, the two distributions, in fact, are the same. This proves that the CNP is a pseudometric on the space of graphs.

\WASSERSTEINTHREE*
\textit{Proof.} Consider the Eigenvalue decomposition of the non-negative, symmetric, real matrix $\Sigma = V \Lambda V^\top$.
In the trace term of the closed form solution of the Wasserstein distance, we have:
\begin{equation*}
    \begin{aligned}
        \operatorname{Tr}(\Sigma_i) &= \operatorname{Tr}(D_i^{-\frac{1}{2}}\Sigma D_i^{-\frac{1}{2}}) = \operatorname{Tr}(D_i^{-\frac{1}{2}} V \Lambda V^\top D_i^{-\frac{1}{2}})\\
        &= \operatorname{Tr}(V^\top D_i^{-\frac{1}{2}} D_i^{-\frac{1}{2}} V \Lambda) = \operatorname{Tr}( D_i^{-1} \Lambda)\\
    \end{aligned}
\end{equation*}
By the same reasoning $\operatorname{Tr}(\Sigma_j) = \operatorname{Tr}( D_j^{-1} \Lambda)$. Regarding the last term, we can use that $V^\top DV = D$ for any diagonal matrix $D$ and the fact that diagonal matrices commute:
\begin{equation*}
    \begin{aligned}
        &(D_i^{-\frac{1}{2}} V \Lambda^{\frac{1}{2}} D_i^{\frac{1}{2}} V^\top D_i^{-\frac{1}{2}})^2 \\
        &= D_i^{-\frac{1}{2}} V \Lambda^{\frac{1}{2}} D_i^{\frac{1}{2}} V^\top D_i^{-\frac{1}{2}} D_i^{-\frac{1}{2}} V \Lambda^{\frac{1}{2}} D_i^{\frac{1}{2}} V^\top D_i^{-\frac{1}{2}}\\
        &= D_i^{-\frac{1}{2}} V \Lambda^{\frac{1}{2}} D_i^{\frac{1}{2}} D_i^{-\frac{1}{2}} D_i^{-\frac{1}{2}} \Lambda^{\frac{1}{2}} D_i^{\frac{1}{2}} V^\top D_i^{-\frac{1}{2}}\\
        &= D_i^{-\frac{1}{2}} V \Lambda V^\top D_i^{-\frac{1}{2}}\\
        &= \Sigma
    \end{aligned}
\end{equation*}
Then the similarly for the last term:
\begin{equation*}
    \begin{aligned}
        &\Sigma_i^{\frac{1}{2}}\Sigma_j \Sigma_i^{\frac{1}{2}}\\
        &= D_i^{-\frac{1}{2}} V \Lambda^{\frac{1}{2}} D_i^{\frac{1}{2}} V^\top D_i^{-\frac{1}{2}} D_j^{-\frac{1}{2}} V \Lambda V^\top D_j^{-\frac{1}{2}} D_i^{-\frac{1}{2}} V \Lambda^{\frac{1}{2}} D_i^{\frac{1}{2}} V^\top D_i^{-\frac{1}{2}}\\
        &= D_i^{-\frac{1}{2}} V \Lambda^{\frac{1}{2}} D_i^{\frac{1}{2}} D_i^{-\frac{1}{2}} D_j^{-\frac{1}{2}}  \Lambda D_j^{-\frac{1}{2}} D_i^{-\frac{1}{2}}  \Lambda^{\frac{1}{2}} D_i^{\frac{1}{2}} V^\top D_i^{-\frac{1}{2}}\\
        &= D_i^{-\frac{1}{2}} V \Lambda^2 D_j^{-1} V^\top D_i^{-\frac{1}{2}}
    \end{aligned}
\end{equation*}
We can now fairly easily see that:
\begin{equation*}
    \begin{aligned}
        (\Sigma_i^{\frac{1}{2}}\Sigma_j \Sigma_i^{\frac{1}{2}})^{\frac{1}{2}} = D_i^{-\frac{1}{2}} V \Lambda D_j^{-\frac{1}{2}} D_i^{\frac{1}{2}} V^\top D_i^{-\frac{1}{2}} 
    \end{aligned}
\end{equation*}
Since the trace is invariant under cyclic permutations, we can write:
\begin{equation*}
    \begin{aligned}
        & \operatorname{Tr}((\Sigma_i^{\frac{1}{2}}\Sigma_j \Sigma_i^{\frac{1}{2}})^{\frac{1}{2}})\\
        &= \operatorname{Tr}(D_i^{-\frac{1}{2}} V \Lambda D_j^{-\frac{1}{2}} D_i^{\frac{1}{2}} V^\top D_i^{-\frac{1}{2}})\\
        &= \operatorname{Tr}(V^\top D_i^{-\frac{1}{2}} D_i^{-\frac{1}{2}} V \Lambda D_j^{-\frac{1}{2}} D_i^{\frac{1}{2}})\\
        &= \operatorname{Tr}(D_i^{-\frac{1}{2}} D_i^{-\frac{1}{2}} \Lambda D_j^{-\frac{1}{2}} D_i^{\frac{1}{2}})\\
        &= \operatorname{Tr}(\Lambda D_j^{-\frac{1}{2}} D_i^{-\frac{1}{2}})\\
        &= \sum_{x=1}^{n} \frac{\lambda_x}{\sqrt{d^{(i)}_x d^{(j)}_x}}
    \end{aligned}
\end{equation*}
Plugging this in yields the claim.
}

\end{document}